\documentclass[onecolumn]{revtex4-2}

\usepackage{graphicx}
\usepackage{amsmath}
\usepackage{amsfonts}
\usepackage{siunitx}
\usepackage[version=4]{mhchem}
\usepackage{caption,subcaption}

\begin{document}

\preprint{APS/123-QED}

\title{Rare Isotope-Containing Diamond Color Centers for Fundamental Symmetry Tests}

\author{I. M. Morris}
\affiliation{Department of Physics and Astronomy, Michigan State University}

\author{K. Klink}
\affiliation{Department of Physics and Astronomy, Michigan State University}

\author{J.T. Singh}
\affiliation{Department of Physics and Astronomy, Michigan State University}

\author{J. L. Mendoza-Cortes}
\affiliation{Department of Chemical Engineering \& Materials Science, Michigan State University}

\author{S. S. Nicley}
\affiliation{Department of Electrical and Computer Engineering, Michigan State University}
\affiliation{Department of Chemical Engineering \& Materials Science, Michigan State University}
\affiliation{Coatings and Diamond Technologies Division, Center Midwest (CMW), Fraunhofer USA Inc.}

\author{J. N. Becker}
\affiliation{Coatings and Diamond Technologies Division, Center Midwest (CMW), Fraunhofer USA Inc.}
\affiliation{Department of Electrical and Computer Engineering, Michigan State University}

\date{\today}

\begin{abstract}
Detecting a non-zero electric dipole moment (EDM) in a particle would unambiguously signify physics beyond the Standard Model. A potential pathway towards this is the detection of a nuclear Schiff moment, the magnitude of which is enhanced by the presence of nuclear octupole deformation. However, due to the low production rate of isotopes featuring such "pear-shaped" nuclei, capturing, detecting, and manipulating them efficiently is a crucial prerequisite. Incorporating them into synthetic diamond optical crystals can produce defects with defined, molecule-like structures and isolated electronic states within the diamond band gap, increasing capture efficiency, enabling repeated probing of even a single atom, and producing narrow optical linewidths. In this study, we used density functional theory (DFT) to investigate the formation, structure, and electronic properties of crystal defects in diamond containing $^{229}Pa$, a rare isotope that is predicted to have an exceptionally strong nuclear octupole deformation. In addition, we identified and studied stable lanthanide-containing defects with similar electronic structures as non-radioactive proxies to aid in experimental methods. Our findings hold promise for the existence of such defects and can contribute to the development of a quantum information processing-inspired toolbox of techniques for studying rare isotopes.
\end{abstract}

\maketitle


\section{Introduction}
The search for charge-conjugation-parity (\emph{CP}) symmetry-violating interactions is a critical aspect of modern physics that aims to answer some of the universe's most fundamental questions. In particular, \emph{CP} violations can help explain the observed baryon asymmetry in the universe \cite{sakharov}. However, current observations of \emph{CP} violations are not significant enough to account for such phenomena. Recently, the measurement of a non-zero permanent electric dipole moment (EDM) within atomic nuclei induced by the nuclear Schiff moment has garnered considerable attention as a potential solution. The existence of a permanent EDM requires breaking of both time-reversal symmetry (\emph{T}) and parity symmetry (\emph{P}), which, by the \emph{CPT} theorem, implies that it breaks \emph{CP} symmetry as well \cite{rmp2019}. Therefore, the study of permanent EDMs in atomic nuclei provides an exciting avenue for detecting \emph{CP}-violating phenomena and addressing some of the most pressing questions in physics today. \par  

Measuring a permanent EDM poses a significant challenge due to its extremely weak signature. However, certain pear-shaped (octupole-deformed) nuclei, such as \ce{^{223}Fr}, \ce{^{225}Ra}, and \ce{^{229}Pa}, have shown to be particularly sensitive to EDM measurements, making them ideal candidates for further study \cite{spevak96,spevak97}. In particular, \ce{^{229}Pa} is predicted to provide over six orders of magnitude more sensitivity than the current experimental limit on EDM measurements taken with \ce{^{199}Hg} \cite{hh83,singh_new_2019}. Despite its potential, the limited global production of \ce{^{229}Pa} has hindered its experimental study. However, the newly-opened Facility for Rare Isotope Beams (FRIB) at Michigan State University is expected to produce a significant amount of \ce{^{229}Pa} within the decade \cite{abel}. This will provide a host of opportunities to study \ce{^{229}Pa}. One such opportunity that has been proposed is to implant \ce{^{229}Pa} nuclei within an optical crystal, thereby enhancing the signal for EDM measurements. This approach provides numerous advantages, such as high number densities, efficient optical probing, and large internal electric fields for oriented non-inversion symmetric crystal defects in optical crystals. However, one other factor that has limited the study of \ce{^{229}Pa} is its extreme toxicity and radioactivity. As such, stable nuclear surrogates are necessary for the development of experimental and testing schemes prior to use of \ce{^{229}Pa}. \ce{^{141}Pr} is an excellent candidate for this as it is expected to be isoelectronic with \ce{^{229}Pa} and has the same nuclear spin $I = \frac{5}{2}$. Moreover, \ce{^{141}Pr} is not as toxic and not radioactive. Thus, \ce{^{141}Pr} can serve as a stable nuclear surrogate to \ce{^{229}Pa}, allowing for method development and testing. This approach can pave the way for future experiments in detecting EDMs in atoms and addressing some of the most profound questions in modern physics. \par

Diamond is a highly suitable host material for EDM-sensitive isotopes such as \ce{^{229}Pa}. It possesses exceptional radiation hardness, making it more resistant to damage from implantation and the decay of incorporated radioactive species than most other host materials \cite{zou2020proton,bauer1995radiation}. Additionally, its wide band gap (5.5 eV) increases the probability of defect formation within the gap, as demonstrated by the existence of thousands of optically active crystal defects in diamond \cite{zaitsev2013optical}. Furthermore, synthetic diamond can be made nuclear spin-free using \ce{^{12}C} enriched precursors in chemical vapor deposition (CVD) growth, eliminating a significant source of spin decoherence and effectively creating an almost perfect spin vacuum \cite{PhysRevMaterials.3.065205}. The extensively studied nitrogen vacancy center in diamond can also be utilized as a highly sensitive quantum magnetometer and can be used for in-situ co-magnetometry \cite{hong2013nanoscale}. Overall, these properties make diamond a highly attractive material for hosting isotopes such as \ce{^{229}Pa} \cite{das2022diamond}. \par

This paper presents a study on the geometric structure, thermodynamic stability, and electronic properties of \ce{^{229}Pa} and \ce{^{141}Pr} defects in diamond using density functional theory (DFT). Specifically, we investigate a variety of different defect configurations, including substitutional defects as well as defects with one to four vacancies introduced nearby. The paper is organized as follows: Section II outlines the computational details and methods involved in the DFT calculations; Section III presents results of these calculations, including geometric structure, formation energies, and charge transition levels along with electronic structure and EDM sensitivity. Finally, in Section IV, we draw conclusions based on our findings. \par

\begin{figure}[htb]
\begin{center}
\includegraphics[width=\textwidth,angle=0]{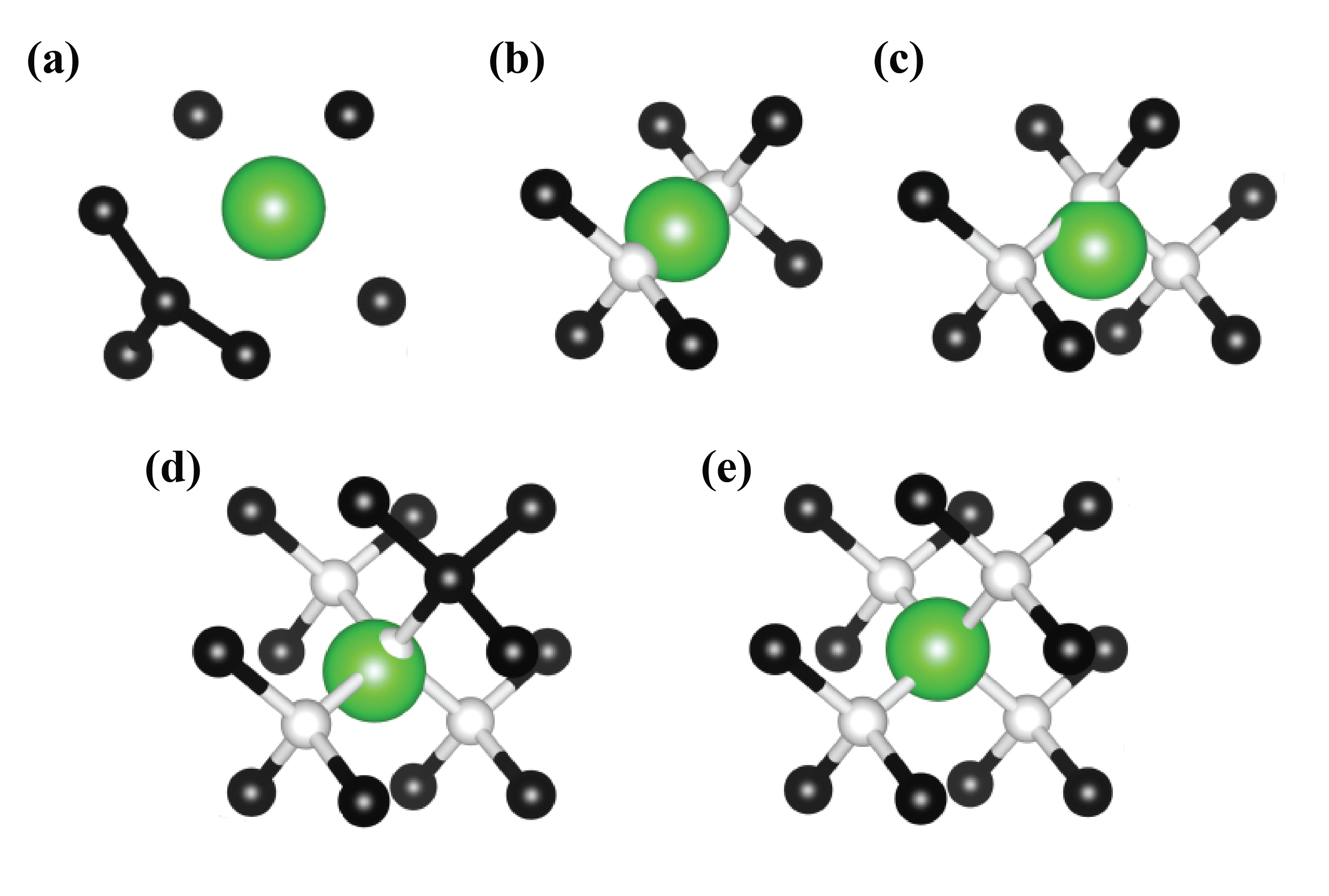}
\caption{Relaxed defect structures of (a) \ce{^{229}Pa}$_{\text{sub}}$, (b) \ce{^{229}PaV}, (c) \ce{^{229}PaV}$_2$, (d) \ce{^{229}PaV}$_3$, and (e) \ce{^{229}PaV}$_4$. For clarity, only the defect ion and the nearest neighbor carbon atoms are displayed. The larger green atom is the $^{229}$Pa ion. The black atoms represent carbon, while the white atoms represent vacancies. For \ce{^{229}PaV}$_2$, \ce{^{229}PaV}$_3$, and \ce{^{229}PaV}$_4$ the "extra" white vacancy ball that can be seen through the \ce{^{229}Pa} is the initial position of the \ce{^{229}Pa} at a lattice site. }
\label{defect_structures}
\end{center}
\end{figure}

\section{Methods}
Spin-polarized DFT was employed using the projector augmented wave method \cite{blochl_projector_1994,bengone_implementation_2000} as implemented in VASP 6.2.1 \cite{kresse_efficient_1996} to characterize isotopic \ce{^{229}Pa} and \ce{^{141}Pr} defects in diamond. The exchange and correlation behavior of the valence electrons ($2s^22p^2$, $6d^27s^25f^1$, and $4f^3s^2$ electrons for C, Pa and Pr, respectively) during structure optimization was described using the Perdew-Burke-Ernzerhof (PBE) generalized gradient approximation (GGA) \cite{perdew_generalized_1996}. To account for the strongly correlated behavior of the \emph{f}-electrons in actinides and lanthanides, a Hubbard-U-type correction (DFT+U) was included for Pa and Pr \emph{f}-electrons in all PBE-level calculations. The implementation suggested by Leichtenstein et al. \cite{beridze_benchmarking_2014} was used with an on-site Coulomb parameter U = 7 eV and on-site exchange parameter J = 1 eV for Pa and Pr, as has been used by others to study lanthanide defects in diamond \cite{vanpoucke_can_2019,tan_study_2020,magyar2014synthesis,tan_structural_2018}.  Additionally, the Heyd-Scuseria-Ernzerhof (HSE06) hybrid functional \cite{heyd_hybrid_2003,krukau_influence_2006} was used for the calculation of highly accurate electronic structures. This range-separated hybrid functional can accurately reproduce experimental band gaps and charge transition levels in diamond and other group-IV semiconductors to within 0.1 eV \cite{deak_accurate_2010,gali_identification_2009} and has successfully described a variety of defects in diamond \cite{vanpoucke_linker_2017,henderson_accurate_2011,garza_predicting_2016,hendrickx_understanding_2015,chaudhry_first-principles_2014,gali_identification_2009,czelej_electronic_2017}.

A variety of defect configurations were studied, including defect ions placed in the substitutional lattice site as well as in substitutional lattice site positions with one to four vacant sites adjacent to them. Calculations were performed on a 3x3x3 diamond supercell containing 216 atoms, and the Brillouin zone was sampled at the $\Gamma$ point. The excited states were calculated using the constrained-occupation DFT method ($\Delta$-SCF) \cite{gali_identification_2009} with zero phonon lines (ZPL) calculated by taking the energy difference between ground and excited states. The initial geometries of the models are depicted in Figure \ref{defect_structures}. The supercell defects were allowed to relax with a constant volume using a conjugate gradient method to ensure that the defect formation energies are comparable. The plane-wave energy cutoff was set to 370 eV. Ionic optimization was performed until forces were less than 10$^2$ eV/\AA{}, and the break condition for the electronic self-consistent loop was set to 10$^6$ eV. To account for the isotopic nature of $^{229}$Pa and $^{141}$Pr, the mass value in the POTCAR file was changed accordingly.\par

The PBE functional was chosen for geometry relaxation due to its lower computational cost and its ability to predict the structures of a variety of defects in diamond with sufficient accuracy \cite{palma_vn2_2020,Davidsson_2018,drumm_thermodynamic_2010,cajzl_erbium_2017}. Furthermore, using a smaller 2x2x2 supercell of 64 atoms, we relaxed the geometry using both PBE+U and HSE06 functionals and found that the difference in atomic positions between the two relaxed structures was less than $10^{-5}$ \AA{} on average, demonstrating that the accuracy of PBE+U is comparable to that of HSE06 for geometry relaxation. \par

To assess which defect configuration was most stable, formation and cohesive energies were calculated for each defect studied. The formation energy for a defect $X$ with charge state $q$ can be calculated according to:
\begin{equation}
E^{\text{f}}[X^q] = E_{\text{tot}}[X^q] - E_{\text{tot}}[\text{bulk}] - \sum_{i=1}^{k} n_i\mu_i + q(\epsilon_{\text{VBM}} + E_{\text{F}}) + E_{\text{corr}},
\end{equation}
where $E_{\text{tot}}[X^q]$ and $E_{\text{tot}}[\text{bulk}]$ are the total energies of the bulk material with and without the defect, respectively; $n_i$ is the number of atoms of species $i$ that have been added to or removed from the supercell (for example, $\ce{^{229}Pa_{\text{sub}}}$ removed 1 C and added 1 \ce{^{229}Pa}); $\mu_i$ is the chemical potential corresponding to atomic species $i$; $\epsilon_{\text{VBM}}$ is the valence band maximum of the bulk material; $E_{\text{F}}$ is the Fermi level, which can have values within the material's band gap; and $E_{\text{corr}}$ is the finite-size electrostatic correction \cite{komsa_finite-size_2012}. $E_{\text{corr}}$ was obtained using the scheme proposed by Freysoldt, Neugebauer and Van de Walle (FNV) \cite{freysoldt_fully_2009} as implemented in the Spinney code package \cite{arrigoni_spinney_2021}. The chemical potential for C was obtained by dividing the total energy of the pristine diamond supercell by the number of atoms. The chemical potential of Pa was calculated as the total energy of metallic Pa (with a bcc tetragonal structure, I4/mmm, no. 139) divided by the number of Pa atoms. Similarly, the chemical potential of Pr was calculated using the total energy of metallic Pr (with a hexagonal structure, P63/mmc, no. 194) divided by the number of Pr atoms. For the chemical potential, the k-point sampling was increased to 9x9x9 due to the small crystal structure.  \par

Cohesive energies were calculated according to:
\begin{equation}
E_{\text{c}} = \frac{1}{n}\left(\sum_{i=1}^{k} n_i E_{\text{atom},i} - E_{\text{tot}}\right),
\end{equation}
where $n$ is the total number of atoms, $E_{\text{tot}}$ is the total energy of the defect system, $n_i$ is the number of atoms of species $i$, and $E_{\text{atom},i}$ is the energy per atom for species $i$ \cite{doi:10.1063/1.4867544}. In order to evaluate the cohesive energies of the structures, it was necessary to calculate the total energies of the corresponding isolated atoms in the structures (using the same exchange functionals and calculation quality settings). For the calculation of the C atom, a 10x10x10 \AA{} cube with a single C atom in the center was used, giving enough space around the atom for it to be considered as an isolated atom. For the Pa and Pr atoms, a slightly larger cube of 15x15x15 \AA{} was used to ensure isolation of the atoms. \par

To determine the most stable charge state for a given defect, its charge transition levels (CTLs) are calculated. The CTL is the Fermi level at which a transition between two charge states becomes energetically favorable \cite{freysoldt_first-principles_2014}. It is calculated using the formula:
\begin{equation}
\epsilon(q_1/q_2) = \frac{E_{q_1}^{\text{tot}} + E_{q_1}^{\text{corr}} - E_{q_2}^{\text{tot}} - E_{q_2}^{\text{corr}}}{q_2 - q_1}
\end{equation}
where $E_{q}^{\text{tot}}$ is the total energy of the supercell calculation in charge state $q$ and $E_{q}^{\text{corr}}$ is the corresponding charge correction that accounts for the periodic interaction of charges between neighboring supercells \cite{sundararaman_first-principles_2017,freysoldt_fully_2009,goyal_conundrum_2019}. \par

The formation and cohesive energies were evaluated for the neutral charge states of the different defect configurations to determine which configuration is most stable. From there, we limited our analysis to the most stable structure and plotted the formation energies for different charge states as a function of Fermi level to determine which charge state is most stable (determined by which charge state has the lowest formation energy for any given Fermi level). The crossing points of these formation energy lines represent charge transition levels, where one charge state becomes more favorable than another. \par 

Zero-field splitting (ZFS), magnetic hyperfine, and electric field gradient tensors were all calculated within VASP. For the ZFS tensor in particular, we use the method by \cite{PhysRevB.90.235205} with the PBE functional, which has been demonstrated to be sufficiently accurate \cite{Davidsson_2018}. For all of these calculations, a higher cut-off energy of 700 eV was used. VESTA \cite{momma2008vesta} was used to visualize the defect structures in addition to the wave functions, whose plane wave coefficients we extracted using the Python class PyVaspwfc \cite{zheng_qijingzhengvaspbandunfolding_nodate}. Similarly, transition dipole moments were calculated directly using said plane wave coefficients. \par

\section{Results and Discussion}
\subsection{Structure and Stability}
First, the structure of each of the defect configurations was studied to determine which is thermodynamically most likely to form during ion implantation and subsequent annealing. All defect ions were initially placed at a substitutional lattice site with nearest neighbor C atoms removed to create vacancies. The final relaxed structures for each defect configuration are shown in Figure \ref{defect_structures}. Interestingly, we find that \ce{^{229}Pa} and \ce{^{141}Pr} defects form qualitatively identical structures for all defect models considered. As such, the following descriptions and images for each defect configuration apply to both. \par

For the substitutional defect with no vacancies, the defect ion did not move, but the nearest neighbor carbon atoms were displaced outwards. For the single vacancy, the ion moved into the split vacancy configuration, while for the higher-order vacancy complexes, it moved into a position that filled the void created by the removed carbons. It is worth noting that while the split vacancy is inversion symmetric, the higher-order vacancy complexes are not, resulting in a permanent electric dipole moment and thus static internal electric field. While this is usually avoided for quantum information processing applications as it makes defects susceptible to environmental field fluctuations, it is in fact desirable for EDM experiments, as it increases the sensitivity of EDM measurements \cite{singh_new_2019}. \par

The calculated formation and cohesive energies shed light on which configuration is most energetically favorable to form. Our analysis reveals that, for both \ce{^{229}Pa} and \ce{^141Pr}, the substitutional model is less favorable than those containing vacancies, as it generally has a higher formation energy despite having a marginally higher cohesive energy in certain cases. This is in agreement with other first principles studies of defects in diamond that feature large ions, which can introduce significant strain \cite{tan_principle_2020,tan_structural_2018,cajzl_erbium_2017,tan_study_2020}. The introduction of vacancies helps to offset this by creating additional room for the dopant atom. Among the defects with vacancies, the double and triple vacancies are the most energetically favorable in terms of formation energy and have comparable or larger cohesive energy values compared to the single and quadruple vacancy defects. Between the double and triple vacancy, however, we find that the double vacancy is the most stable, as there is diminishing gains by adding yet another vacancy \cite{vanpoucke_can_2019}. Moreover, higher-order vacancy complexes are kinetically less likely to form due to the low mobility of substitutional defects in diamond at typical processing temperatures \cite{mainwood_nitrogen_1994}. Additionally, a similar ab-initio study was done on Ce defects in diamond, and CeV$_2$ was found to be most stable \cite{tan_study_2020,tan2021first}. Therefore, we conclude that the most stable structure for both \ce{^{229}Pa} and \ce{^141Pr} defects in diamond is a defect ion accompanied by two vacancies. \par

To further assess the probability of these defects forming, we compare our calculated results with values calculated for other defects in diamond. A combined experimental and theoretical study done on Er$^{3+}$ ions implanted in diamond calculated similar cohesive and formation energies. Importantly, they also experimentally observed the characteristic telecom band emission from the Erbium ions after implantation and annealing \cite{cajzl_erbium_2017}. Additionally, formation energies calculated for nickel complexes in diamond also find similar formation energies. Lastly, we compare the formation energies of the various charge states with the common NV center (see Figure 4) and they are the same order of magnitude. These examples demonstrate the feasibility of lanthanides and actinides forming luminescent centers in diamond. \par

Notably, non-inversion-symmetric defect configurations are preferred. As noted above, while typically not ideal for quantum information processing applications, \cite{rose2018observation}, the opposite is true for EDM measurements \cite{nayak2006ab,fiebig2016evolution}. As was stated above, this results in a permanent electric dipole moment which can result in linear Stark shifts. This results in symmetric doublet splittings in optical transitions when two states with oppositely-oriented dipole moments are degenerate in a non-inversion symmetric site. This makes it possible to specifically address ions that have a specific direction of electric polarization. Going back to the initial statement of inversion-breaking defects being preferred, this means that for any given sample, the majority of defects formed will not be inversion-symmetric and will thus retain the advantages of enhanced EDM measurement sensitivity. Based on these findings, we focused subsequent calculations on \ce{^{229}Pa} and \ce{^141Pr} defects with two vacancies. \par

\begin{figure}[htb]
\begin{center}
\includegraphics[width=\textwidth]{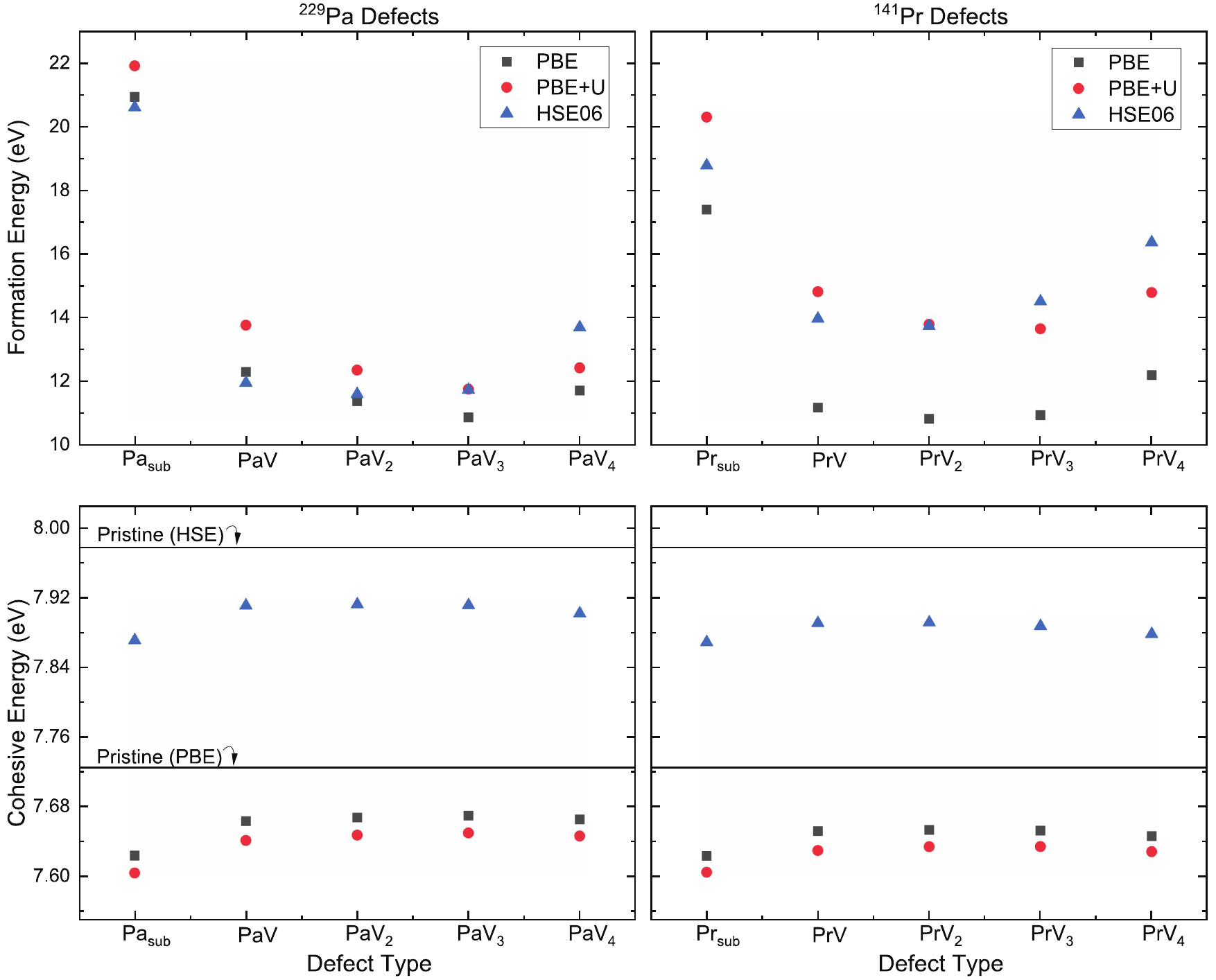}
\caption{Top panels are formation energies for different defect configurations using different functionals. Bottom panels are cohesive energies for different defect configurations and different functionals. For the bottom panels, the solid lines denote the cohesive energy for pristine diamond without any defects using both PBE and HSE06 functionals.}
\label{formation_vs_fermi_energy}
\end{center}
\end{figure}

\subsection{Charge State Formation Energies}
Since $^{\text{229}}$PaV$_{\text{2}}$ and $^{\text{141}}$PrV$_{\text{2}}$ appear to be the most thermodynamically favorable defects, which also feature promising geometries for EDM measurements, we will focus on these configurations from here on out. We start by determining their potential charge states and charge transition levels. Our findings indicate that both defects can potentially take on charge states ranging from -3 to +1.  The formation energy as a function of Fermi level is displayed in Figure \ref{formation_vs_fermi_energy}. As mentioned above, the calculated formation energy values are comparable to those of other defects in diamond that contain large defect ions \cite{cajzl_erbium_2017,vanpoucke_can_2019}, indicating that, at least thermodynamically, defect formation is possible. Additionally, for the purposes of NV co-magnetometry, the -1, -2, and -3 charge states for both defects all land squarely within the Fermi levels where negatively charged NV centers are likely to form. Furthermore, since NV co-magnetometry requires a relatively high donor concentration, the negative charge states are preferred. In terms of which charge state is most likely to form in diamond without the need careful doping, the -1 charge state is nearest the Fermi level for natural   \par

Charge transition levels (CTLs) were calculated using the information shown in the formation energy diagram (Figure \ref{formation_vs_fermi_energy}). Calculating the CTLs for different doping levels allowed us to determine which charge state is most stable at each Fermi level. This information is important for understanding the behavior of the defects in diamond under different doping conditions. For example, it provides information on the effect of different atomic species on the defects' charge stability. Here, the negative charge states act as electron acceptors and require compensatory electron donors in the system, such as substitutional N that have a deep donor level located 1.7 eV below the conduction band minimum \cite{mainwood_substitutional_1979,kobashi2014shift}. \par

\begin{figure}[htb]
\begin{center}
\includegraphics[width=\textwidth,angle=0]{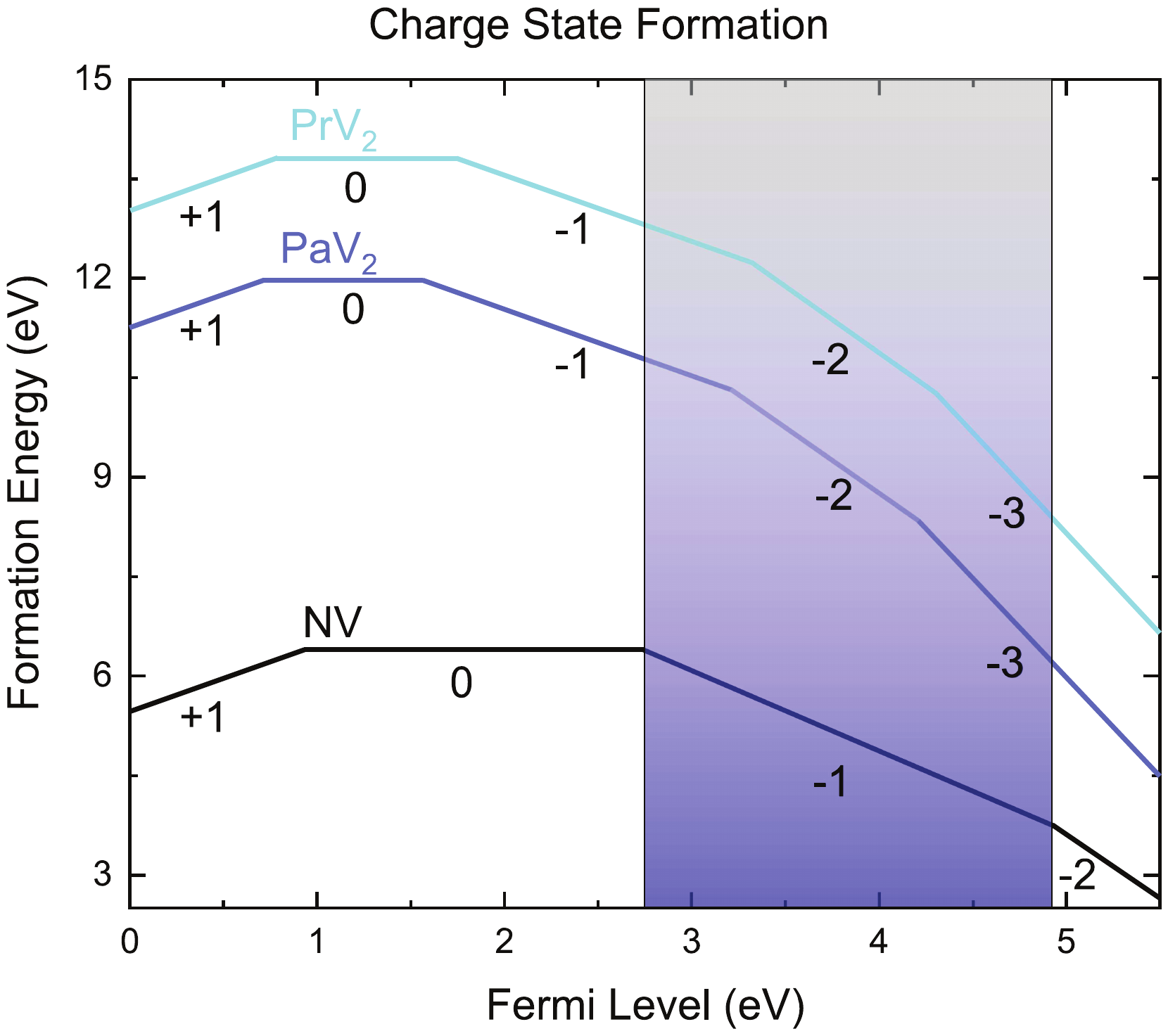}
\caption{Charged formation energy as a function of Fermi level for $\text{PaV}_2$ and $\text{PrV}_2$ defects in diamond. Additionally, formation energies calculated by \cite{Gali+2019+1907+1943} for the NV center in diamond are included for comparison.}
\label{charged_formation_energies}
\end{center}
\end{figure}

\subsection{Electronic Structure}
\begin{figure}[htb]
\begin{center}
\includegraphics[width=\textwidth,angle=0]{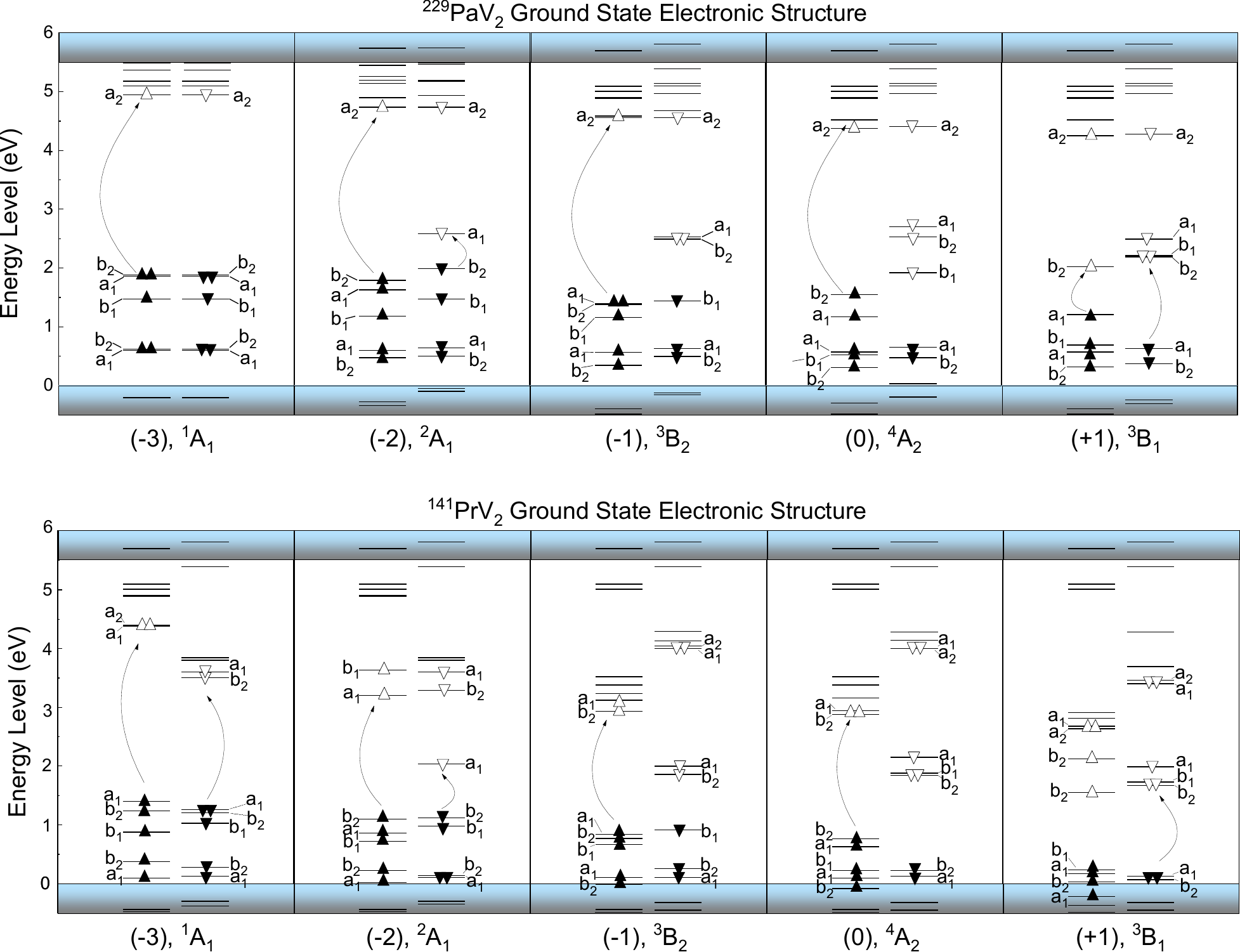}

\caption{Ground state electronic structure for the charge states of $\ce{^{229}PaV}_2$ and $\ce{^{141}PrV}_2$. The single-electron orbitals are labeled with their corresponding irreducible representations. }
\label{elec_struc}
\end{center}
\end{figure}

In this section, we present a detailed analysis of the electronic structure of the defects using group theory and DFT calculations. Both defects of interest, $\ce{^{141}Pr}V_2$ and $\ce{^{229}Pa}V_2$, are part of the C$_{2\text{v}}$ symmetry point group. Using this, we derive a defect molecular orbital diagram to make predictions for the optical transitions and fine structure. First, we calculate the spin-polarized level structure of the single-electron orbitals using DFT and derive which irreducible representation of C$_{2\text{v}}$ they belong to by applying the respective symmetry operators to the calculated wave functions. 
Figure \ref{elec_struc} shows the visualizations of the defect wave functions obtained from the DFT calculations, along with the single-electron orbital levels. From these single electron orbitals, we construct the many electron (molecular) orbital configurations shown in Figure 4. This diagram displays the single-electron Kohn-Sham energy levels and their corresponding irreducible representations, allowing  possible optical transitions to be identified. From the results, the +1 and -1 charge states features S=1 spin triplets; the neutral charge state has a S=3/2 quartet; and the -2 and -3 charge states feature an S=1/2 doublet and S=0 singlet, respectively. With this, we can identify charge states of interest based on their spin. To detect a nuclear-Schiff moment, there needs to be hyperfine coupling to an electron spin, ruling out the -3 charge state as a potential candidate as there is no electron for the nuclear spin to couple to. That leaves the positive and neutral as well as remaining two negative charge states. We focus on the -1 and -2 charge states as they are more likely to form within natural diamond while also falling within Fermi level regions that the negatively charged NV center does. \par

One additional observation from the Kohn-Sham orbitals is that the defect ions introduce occupied bands within the band gap of diamond. This differs from other color centers in diamond such as the group-IV vacancies or nickel vacancies where the Kohn-Sham orbitals are situated below the valence band edge \cite{PhysRevX.8.021063,thiering_eg_2019}. For $\ce{^{229}PaV_2}$ and $\ce{^{141}PrV_2}$, however, both the ground and excited state levels for the minority spin channel are located within the band gap. This localization of the defect states from the bulk bands reduces the probability of single-photon transitions from the defect to bulk states during defect-defect transitions, potentially enhancing the excitation efficiency for the studied centers. \par

In general, our goal is to identify electronic structures featuring ground and excited state spin and orbital configurations leading to qualitatively identical interaction, splitting patterns and optical transitions for both $\ce{^{141}PrV}_2$ and $\ce{^{229}PaV}_2$ so that the stable $\ce{^{141}PrV}_2$ defect can be used as a test bed for method development. To identify these transitions, we first need to know which optical transitions are electric dipole allowed. In C$_{2\text{v}}$, the dipole moment vector transforms as $(\text{B}_1, \text{B}_2, \text{A}_1)$. With that, we can calculate the matrix elements for various transitions using the irreducible symmetry representation to determine which transitions are allowed. Because the electric dipole operator does not act on the spin part of the wavefunction, we only consider non-spin flipping excitations within the spin up and spin down channels, respectively, to identify possible optically allowed transitions. Based on ground and excited state wavefunctions, we calculate transition dipole moments to determine the strength of certain transitions. Using this, transitions which match for both $\ce{^{141}PrV}_2$ and $\ce{^{229}PaV}_2$ and which have a TDM $>$ 1 Debye were selected for further study. The results are displayed in Table \ref{table:tdm_zpl} along with the calculated ZPL for each transition. \par

\begin{table}[!htt]
\begin{center}
\def\arraystretch{1.05}%
\begin{tabular}{c|cccc} \toprule
    {} & {Spin Channel} & {Transition} & {ZPL} & {TDM} \\ \toprule
    {} & {} & {} & {(nm)} & {(Debye)} \\
    {$\ce{^{229}PaV}_2^{2-}$} & {up} & {$^2$A$_1$ $\rightarrow$ $^2$B$_2$} & {533} & {1.37} \\
    {$\ce{^{141}PrV}_2^{2-}$} & {up} & {$^2$A$_1$ $\rightarrow$ $^2$B$_2$} & {764} & {1.61}\\ 
    {} & {down} & {$^2$A$_1$ $\rightarrow$ $^2$B$_2$} & {1765} & {1.16} \\ 
    {$\ce{^{229}PaV}_2^{1-}$} & {down} & {$^3$B$_2$ $\rightarrow$ $^3$A$_2$} & {1187} & {6.64} \\
    {$\ce{^{141}PrV}_2^{1-}$} & {down} & {$^3$B$_2$ $\rightarrow$ $^3$A$_2$} & {1259} & {7.72} \\  
 
\end{tabular}
\caption{Matching optical transitions for the $-2$ and $-1$ charge states for $\ce{^{229}Pa}V_2$ and $\ce{^{141}Pr}V_2$ defects. Spin channel refers to whether a spin up or down electron was promoted to a higher band (i.e., excited state). Transition shows the symmetry of the ground and excited state. TDM is the transition dipole moment, which corresponds to the strength of the transition. |$\Delta D$| is the magnitude of the difference in induced dipole moment from ground to excited state. }
\label{table:tdm_zpl}
\end{center}
\end{table}

We start our analysis with the -1 charge states for both defects, which have $^3\text{B}_2 \rightarrow ^3\text{A}_2$ transitions. These states transform as orbital singlets, due to the fact that the defects have C$_{2\text{v}}$ symmetry, so there is no spin-orbit or Jahn-Teller contributions to the ground or excited states. The ground state $^3\text{B}_2$ features an $S = 1$ spin triplet, and $^{229}$Pa nuclear spin $I = \frac{5}{2}$, leading to the fine and hyperfine structure shown in Figure \ref{pav2-_optical_trans}. Five interactions were considered when analyzing the possible spin state splittings: the electron-electron magnetic dipolar interaction \textbf{D}, the hyperfine interaction \textbf{A}, the nuclear quadrupole interaction \textbf{Q}, and the electronic and nuclear Zeeman interactions: 
\begin{equation}
    \hat{H} = \mu_{B} g_e \vec{B} \cdot \vec{S} + \mu_{N} g_n \vec{B} \cdot \vec{I} + \vec{S}\cdot\bold{D}\cdot \vec{S} + \vec{S}\cdot\bold{A}\cdot\vec{I} + \vec{I}\cdot\bold{Q}\cdot\vec{I}, 
\end{equation}
where $\mu_{B}$ and $\mu_{N}$ are the Bohr and nuclear magnetons, $\vec{B}$ is the magnetic field vector, $g_e$ and $g_n$ are the electron and nuclear g-factors, and $\vec{S}$ and $\vec{I}$ are the total electron and nuclear spin angular momenta. \par
In the principle axis coordinates, the latter three terms in Eq. (3.1) can be converted to:
\begin{equation}
    \hat{H}_{D} = D[S_z^2 - \frac{1}{3}S(S+1) + \frac{\epsilon}{3}(S_+^2 + S_-^2)]
\end{equation}
\begin{equation}
    \hat{H}_{Q} = \frac{eQ_IV_{zz}}{4I(2I-1)}[3I_z^2 - I(I+1) + \eta(I_x^2 + I_y-^2)]
\end{equation}
\begin{equation}
    \hat{H}_{A} = A_{zz}S_zI_z + A_{xx}S_xI_x + A_{yy}S_yI_y 
\end{equation}
where $D = \frac{3}{2}D_{zz}$, $Q_I$ is the nuclear quadrupole moment, $e$ is the electric charge, and $\epsilon = (D_{xx} - D_{yy})/D_{zz}$ and $\eta = (V_{xx} - V_{yy})/V_{zz}$ are the asymmetric coefficients. It should be noted that VASP has been found to underestimate the zero-field tensor, \textbf{D}, even when using hybrid functionals, so the values may be larger than what was calculated \cite{zemla2020graphene}. The quadrupole for $^{229}$Pa has not been experimentally measured, so the theoretically calculated value from \cite{flambaum2022enhanced} was used. The calculated values are shown in Table \ref{table:symmetry_etc} for the defect ions of interest as well as for the nearest neighbor carbons which surround the defects. The calculation can be simplified by selecting the $C_2$ axis of the defects as the z-direction and accounting for the fact that $\hat{H}_D \gg \hat{H}_A, \hat{H}_Q$. This reduces the effective Hamiltonian to 
\begin{equation}
    \hat{H} = \mu_{B} g_eB_zS_z + \mu_{N} g_nB_zI_z + DS_z^2 + QI_z^2 + A_{zz}S_zI_z.
\end{equation}

For the level structure, this means that there is an initial zero-field splitting originating from \textbf{D} which splits the $m_s = 0$ and $m_s = \pm 1$ states. Then, the electron Zeeman interaction lifts the degeneracy of the $m_s = \pm 1$ states. The quadrupole interaction splits each of these branches further into $m_I = \pm \frac{1}{2}, \pm \frac{3}{2}, \pm \frac{5}{2}$ levels. Then, the hyperfine and nuclear Zeeman terms lift the degeneracy of these levels. Similarly, the excited state has the same fundamental splitting pattern. \par

\begin{figure}[htb]
\begin{center}
\includegraphics[width=1\textwidth,angle=0]{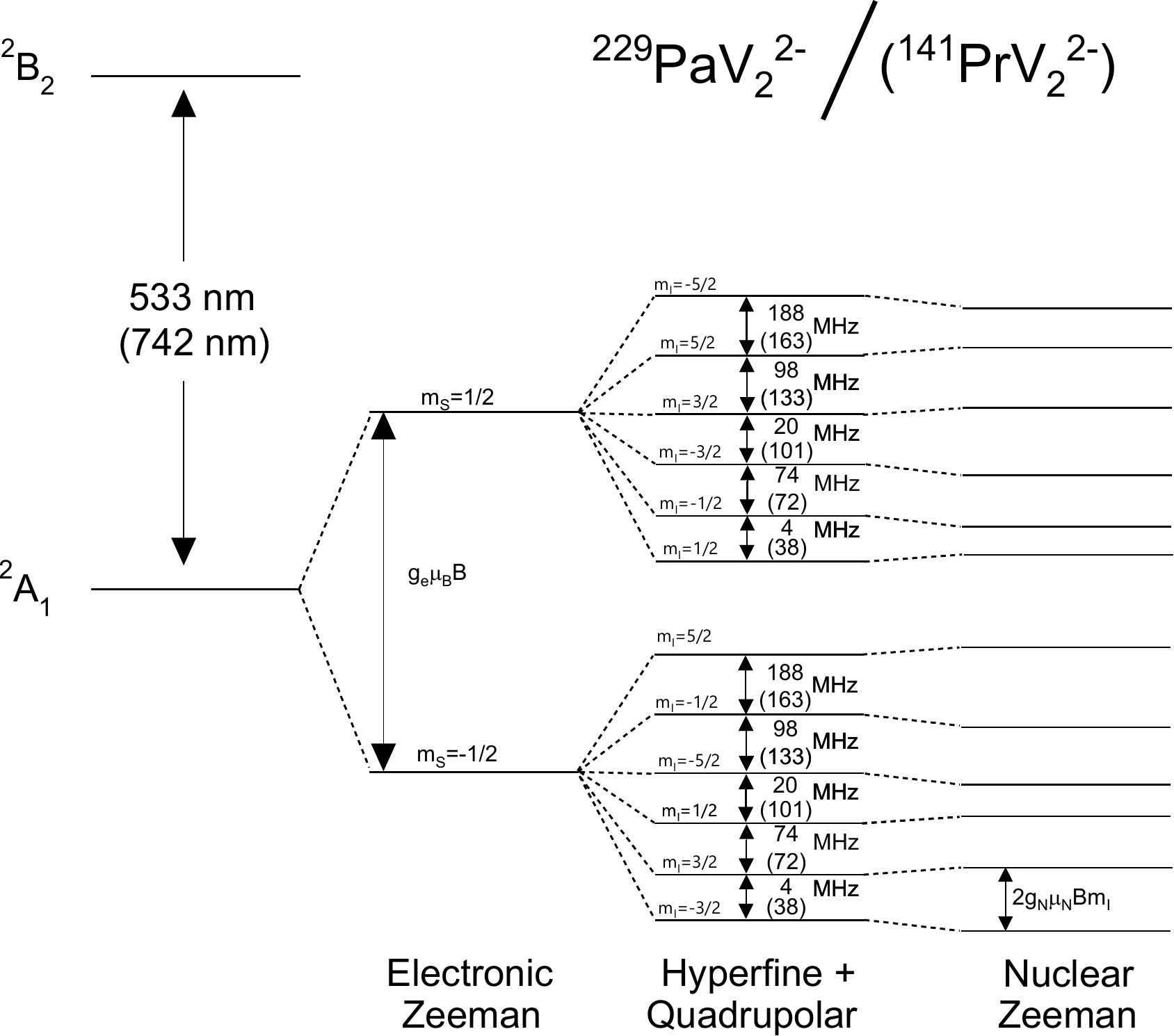}
\caption{Electronic level structure and optical transition of $\ce{^{229}PaV}_2^{2-}$ and $\ce{^{141}PrV}_2^{2-}$.  }
\label{pav1-_optical_trans}
\end{center}
\end{figure}

\begin{figure}[htb]
\begin{center}
\includegraphics[width=1\textwidth,angle=0]{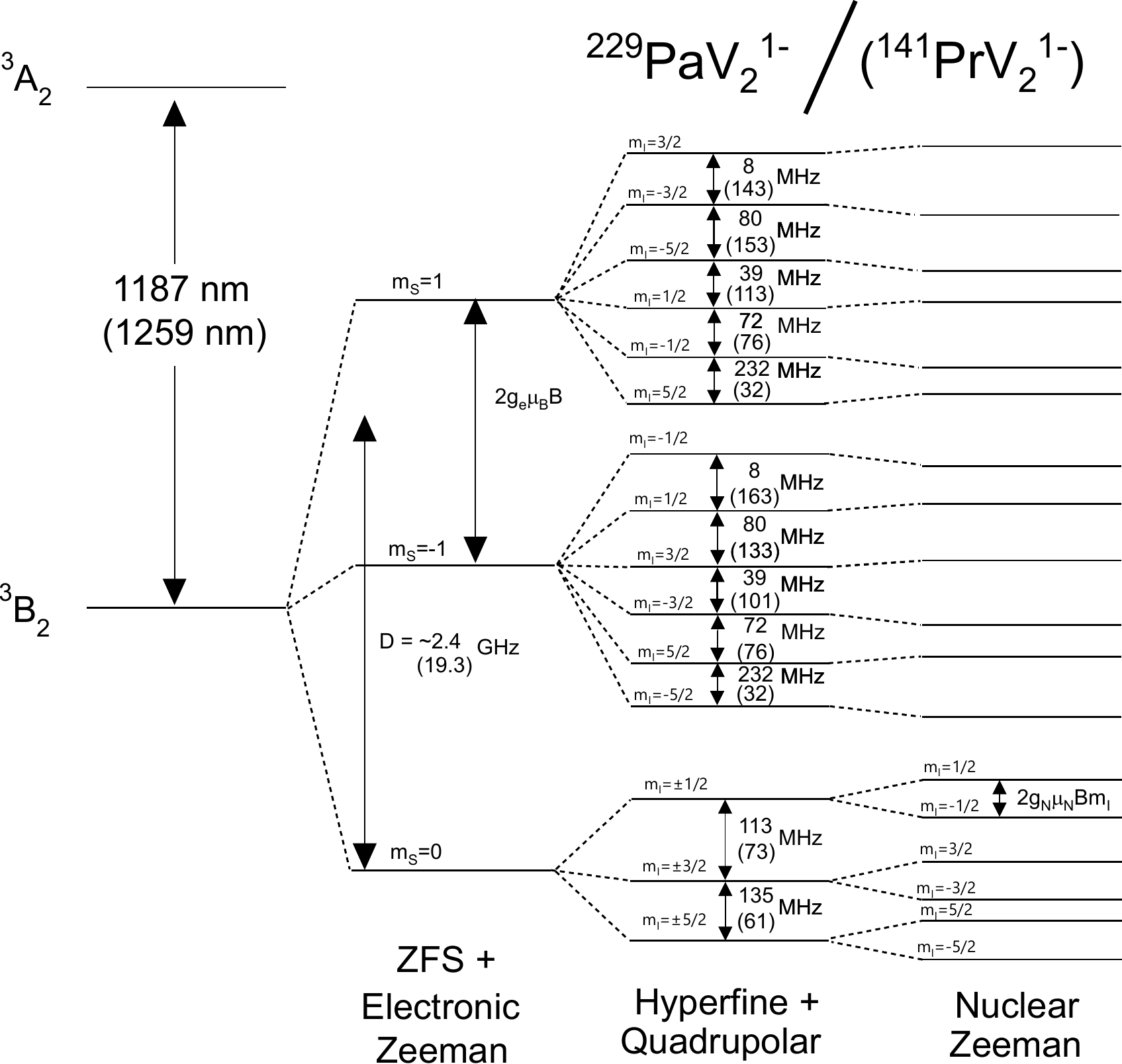}
\caption{Electronic level structure and optical transition of $\ce{^{229}PaV}_2^{1-}$ and $\ce{^{141}PrV}_2^{1-}$.}
\label{pav2-_optical_trans}
\end{center}
\end{figure}

The -2 charge state has one unpaired electron and therefore has spin $S = \frac{1}{2}$ due to the half-occupied molecular orbital that can be spin up or down, resulting in a spin multiplicity of 2. The ground state has the electronic configuration [$a_1$]$^2$[$b_2$]$^2$[$b_1$]$^2$[$a_1$]$^1$[$b_2$]$^2$, which transforms as the irreducible representation A$_2$ based on the direct product of the irreducible representations that constitute the state. Similar to the -1 transitions, the excited and ground states are orbital singlets, resulting in the absence of Jahn-Teller instability and spin-orbit coupling. Similarly, there is no spin-spin interaction because there is only one unpaired electron. Thus, we start our analysis with the application of a magnetic field, which lifts the degeneracy of the $m_s = \pm \frac{1}{2}$ states for both the ground and excited states, as they are both spin doublets. The electronic Zeeman interaction term was given above. Similar to the -1 charge state, there are 6 splittings for each branch from the nuclear Zeeman and magnetic hyperfine interaction terms, which split the $m_I = \pm \frac{1}{2}, \pm \frac{3}{2}, \pm \frac{5}{2}$ levels. See Figure \ref{pav1-_optical_trans} for the full level structure along with calculated values. \par

\begin{table}[!ht]
\begin{center}
\def\arraystretch{1.2}%
\begin{tabular}{c|cccccccc} \toprule
    {} & {Symmetry} & {$A_{xx}$} & {$A_{yy}$ } & {$A_{zz}$ } & {$V_{zz}$ } & {$\eta$} & {$D_{zz}$} & {$\epsilon$} \\ 
    {$\ce{^{229}PaV$_2$}^{2-}$} & {$^2$A$_1$} & {-50.874} & {-48.39} & {-116.319} & {613.191} & {0.595} & {--} & {--} \\
    {$\ce{^{141}PrV$_2$}^{2-}$} & {$^2$A$_1$} & {-195.352} & {-160.946} & {-201.971} & {287.623} & {0.744} & {--} & {--} \\
    {$\ce{^{229}PaV$_2$}^{1-}$} & {$^3$B$_2$} & {-76.684} & {-46.146} & {-88.272}& {-683.679} & {0.776} & {1.611} & {0.115}  \\
    {$\ce{^{141}PrV$_2$}^{1-}$} & {$^3$B$_2$} & {-106.872} & {-104.617} & {-113.123} & {370.684} & {0.77} & {12.931} & {0.275} \\
     \end{tabular}
     \caption{Values for the zero-field splitting (D$_{zz}$) in GHz (for S=1), hyperfine coupling parameters (A$_{xx}$, A$_{yy}$, A$_{zz}$) in MHz, electric field gradients (V$_{zz}$) in V/\AA{}$^2$, and symmetry labels for the ground state of each defect in the -2 and -1 charge states, all of which were calculated using HSE06 aside from the zero field splitting tensor which used the PBE functional.}
     \label{table:symmetry_etc}
\end{center}
\end{table}

\subsection{Estimated EDM Sensitivity}
To provide a first estimate of the sensitivity of these defects for EDM measurements, we focus on the effective internal field generated by their non-inversion symmetric structure. To do this, we first attempt to estimate the differential dipole moment, $\Delta\mu$, and polarizability, $\Delta\alpha$, between the ground and excited states. We take an approach proposed by \cite{bathen2020first,ramachandran2023nuclear} where the Stark shift of the ZPL can be modeled by:
\begin{equation}
    \Delta E_{\text{ZPL}} = -\frac{1}{\epsilon_s}\Delta\mu E - \frac{1}{2\epsilon_s^2}\Delta\alpha E^2, 
\end{equation}
where E is an applied electric field perturbing the defect, and $\epsilon_s$ is the static dielectric constant of the material which we take as 5.7 for diamond \cite{spear1994synthetic}. Using this, we calculate the ZPL at varying applied electric field strengths and fit the equation above to the resulting ZPL energy changes. We performed these calculations for both the -1 and -2 charge state for the $^{229}$PaV$_2$ defects. The VASP applied electric field is in units of eV/\AA{}, so the resulting $\Delta\mu$ is initially in units of e\AA{} and $\Delta\alpha$ is in units of $\text{m}^2\text{e}$/V. We also provide these values in the other typically quoted units of Debye for differential dipole moment and $\text{a}_0^3$ for differential polarizability (see Table \ref{table:effective_e}). \par

With these values in hand, we can make an estimate for the internal effective electric field within the crystal. To do this, we use the following equation: 
\begin{equation}
    E_{\text{eff}} = \left(\frac{1}{4\pi\epsilon_0}\right) \left( \frac{\Delta\mu}{Z_{\text{scale}}^3}\right),
\end{equation}
where, $\epsilon_0 = 8.85 \times 10^{-12}$ [F/m] is the permittivity of free space, and $Z_{scale} = 1 $\AA{} is a length scale estimate for the volume. Importantly, the dipole moments used in the equation for differential dipole moment are not the EDM that is measured for fundamental symmetry breaking. It is the overall induced dipole because of the effective electric field within the non-inversion symmetric defect. With this, we can obtain a rough estimate for the effective electric field experienced by an electron within the defect. This value is then further reduced by a conservative factor of 100 to estimate the shielding by the crystal, also known as the permittivity tensor.

\begin{table}[!ht]
\begin{center}
\def\arraystretch{1.2}%
\begin{tabular}{c|ccccc} \toprule
    {} & {$\Delta\mu$ (e\AA{})} & {$\Delta\mu$ (D)} & {$\Delta\alpha$ ($\text{\AA{}}^2\text{e}$/V)} & {$\Delta\alpha$ ($\text{a}_0^3$)} & {E$_{\text{eff}}$ (GV/cm)}  \\ 
    {$\ce{^{229}PaV$_2$}^{2-}$} & {1.23} & {5.90} & {0.12} & {11.6} & {1.7} \\
    {$\ce{^{229}PaV$_2$}^{1-}$} & {0.02}& {0.095} & {0.025} & {2.43} & {0.028}  
     \end{tabular}
     \caption{Differential dipole moment and polarizabilities in a couple different unit systems (those from VASP, and those typically quoted).}
     \label{table:effective_e}
\end{center}
\end{table}

Outside of these values for an effective field and differential polarizability, the defects have several other advantages. One such advantage is that the angular momentum is zero, so this greatly reduces their coupling to the lattice, enabling narrow optical linewidths. Furthermore, in the case of the -1 charge state, it features a spin-1 triplet, which should extend its coherence time because there will be limited coupling to the spin-1/2 bath within diamond, similar to the NV center. \par

\section{Conclusion}
We have identified the $\ce{^{229}PaV_2}$ defect in diamond as a promising candidate for tests of fundamental symmetry violations. It lacks inversion symmetry, which allows for heightened EDM sensitivity and can also inhabit a number of negatively-charged states, which have similar Fermi levels to the NV center, enabling co-magnetometry with NV centers. Multiple optical transitions which can be captured with laser spectroscopy techniques were identified. Furthermore, a large effective electric field was calculated. Moreover, while production of $\ce{^{229}Pa}$ will occur at the Facility for Rare Isotope Beams, we have also identified a stable lanthanide-containing defect in the form of $\ce{^{141}PrV_2}$ defects in diamond, for which we have identified ground to excited state configurations and transitions that are qualitatively identical to those of the $\ce{^{229}PaV_2}$. This will facilitate experimental method development. While not considered here, the effect of applied electric fields or strain could also serve to enhance the dipole moment and the Hamiltonian terms should be be explored in the future. 

\section{Acknowledgements}
IMM acknowledges support from an Alfred J. and Ruth Zeits Research Fellowship at MSU. KK acknowledges support from a Hantel Endowed Fellowship at MSU. JNB acknowledges support by the Cowen Family Endowment at MSU.
\bibliography{pav2}

\end{document}